\documentclass{appolbcjm}
\usepackage{graphicx,bm}

\newcommand{\pvec}{\bm{p}}
\newcommand{\beq}{\begin{equation}}
\newcommand{\eeq}{\end{equation}}
\newcommand{\beqs}{\begin{eqnarray}}
\newcommand{\eeqs}{\end{eqnarray}}
\newcommand{\bitem}{\begin{itemize}\item}
\newcommand{\eitem}{\end{itemize}}

\newcommand{\ket}[1]{\vert #1\rangle}

\newcommand{\me}[3]{\langle #1\vert\ #2\ \vert #3\rangle}

\newcommand{\wvec}{\mathbf{w}}

\newcommand{\xvec}{\mathbf{x}}

\newcommand{\qqa}{\bm{q}^2_{{\rm cm},a}}
\newcommand{\Bp}{B^{(\Pvec)}}
\newcommand{\Kt}{\widetilde{K}}
\newcommand{\Ktinv}{\widetilde{K}^{-1}}
\newcommand{\Cb}{C_B^{(\Pvec)}}

\newcommand{\St}{\widetilde{S}}
\newcommand{\Stinv}{\widetilde{S}^{-1}}

\newcommand{\dvec}{\bm{d}}
\newcommand{\Pvec}{\bm{P}}
\newcommand{\svec}{\bm{s}}
\newcommand{\nn}{\nonumber}

\newcommand{\Ecm}{E_{\rm cm}}
\newcommand{\nvec}{\bm{n}}
\newcommand{\zvec}{\bm{z}}

\newcommand{\gbw}{g_{\Delta,{\rm BW}}}

\renewcommand{\headerauthor}{C.J.~Morningstar}

\begin{document}
\title{Low-lying baryon resonances from lattice QCD}
\author{Colin Morningstar
\address{Department of Physics, Carnegie Mellon University, \\
Pittsburgh, Pennsylvania 15213, USA}}
\date{October 8, 2025}
\maketitle
\begin{abstract}
Calculating the properties of baryon resonances from quantum chromodynamics
requires evaluating the temporal correlations between hadronic operators 
using integrations over field configurations weighted by a phase associated
with the action. By formulating quantum chromodynamics on a space-time lattice
in imaginary time, such integrations can be carried out non-perturbatively
using a Markov-chain Monte Carlo method with importance sampling.  The energies
of stationary states in the finite volume of the lattice can be extracted
from the temporal correlations.  A quantization condition involving the 
scattering $K$-matrix and a complicated ``box matrix'' also yields a 
finite-volume energy spectrum. By appropriately parametrizing the scattering
$K$-matrix, the best-fit values of the $K$-matrix parameters are those that
produce a finite-volume spectrum which most closely matches that obtained from the
Monte Carlo computations.  Results for the $\Delta$ resonance are presented,
and a study of scattering for energies near the $\Lambda(1405)$ resonance is
outlined, showing a two pole structure.  The prospects for applying this
methodology to the Roper resonance are discussed.
\end{abstract}
  
\section{Introduction}
The interactions between quarks and gluons are described by 
quantum chromodynamics\cite{Fritzsch:1973pi}
(QCD), a quantum field theory based on a non-Abelian local $SU(3)$ gauge symmetry.  
Quarks and gluons are not directly observable, but they bind to form composite 
particles known as hadrons, such as protons, neutrons, and pions, which can be
observed in experiments.  The majority of hadrons are scattering resonances,
unstable short-lived particles.  The famous Roper 
resonance\cite{Roper:1964zza} is the lightest excitation of the 
proton.  

The interactions between hadrons and the formation of hadron resonances
are also described by QCD.  Unfortunately, extracting the properties of bound 
states and resonances is difficult in relativistic quantum field theories,
and the strong coupling nature of QCD much exacerbates this difficulty.  
Calculating the properties of hadron resonances from quantum chromodynamics
requires evaluating the temporal correlations between hadronic operators 
using integrations over field configurations weighted by a phase associated
with the action. Standard perturbative techniques which work well in quantum 
electrodynamics for performing such integrals are nearly useless for the
integrals which must be evaluated in QCD. By formulating QCD on a space-time
lattice in imaginary time, such integrations can be carried out non-perturbatively
using a Markov-chain Monte Carlo method with importance sampling.  The use of
such techniques is broadly known as lattice QCD.

In recent years, lattice QCD has advanced to the point where it can now 
determine the masses and decay widths of unstable hadronic resonances, 
such as the $\rho$ and $\Lambda(1405)$. These calculations begin by evaluating 
the finite-volume energy levels of the multi-hadron states into which the 
resonances decay, using Markov-chain Monte Carlo techniques for the integration
over the fields. Next, parametrized models of the scattering amplitudes are 
constructed and inserted into a well-established quantization condition. 
This condition involves the scattering $K$-matrix and a complicated ``box 
matrix", which together produce a finite-volume spectrum that depends 
on the scattering parameters. By adjusting these parameters to best reproduce 
the energy levels obtained from lattice QCD, the properties of the resonances 
can be extracted from the resulting scattering amplitudes.

A critical component of these calculations is determining the energies of 
stationary states in a finite volume, particularly those involving multi-hadron 
contributions. These energies are obtained from Monte Carlo estimates of time-dependent 
correlation functions built from carefully designed quantum field operators that 
generate the desired states. To compute these correlators, quark propagators from 
various lattice source points must be contracted together. These propagators are inverses 
of extremely large matrices, but only their products with specific source vectors 
are required. For single-hadron operators, translational symmetry allows the use 
of a limited number of source sites. However, multi-hadron operators necessitate 
the inclusion of all spatial sites on a given source time slice, making
reliable estimates difficult. This challenge has been overcome with new methods, 
such as Laplacian Heaviside (LapH) quark-field smearing. LapH smearing reduces the 
computational burden by projecting the quark propagators into a smaller subspace 
defined by eigenvectors of the gauge-covariant Laplacian, enabling feasible use 
of all spatial sites.

This report showcases some our recent lattice QCD results on baryon resonances. The 
results from a study of the $\Delta$ resonance are presented, and an investigation
into the two-pole structure near the $\Lambda(1405)$ is described.  The feasibility
of studying the famous Roper resonance in the near future is discussed.

\section{Outline of Methodology}

To study resonance and scattering properties in lattice QCD, one first
evaluates the finite-volume energies of stationary states corresponding to
the relevant decay products for a variety of total momenta and symmetry 
representations.  The first step in determining these stationary-state 
energies is to evaluate an $N\times N$ Hermitian matrix of
temporal correlations 
   $C_{ij}(t)
   = \langle 0\vert\, O_i(t+t_0)\, \overline{O}_j(t_0)\ \vert 0\rangle
   $
for each total momentum and symmetry representation.  These correlations
involve judiciously designed operators $O_j(t)=O_j[\overline{\psi},\psi,U]$ 
comprised of quark $\psi,\overline{\psi}$ and gluon $U$ field operators which create
the states of interest.  Each temporal correlator can be expressed as a ratio of 
path integrals over the fields
\beq
  C_{ij}(t)= \frac{ \int {\cal D}(\overline{\psi},\psi,U)\ \ 
   O_i(t+t_0)\ \overline{O}_j(t_0)\ \ \exp\left(-S[\overline{\psi},\psi,U]\right)}{  
  \int {\cal D}(\overline{\psi},\psi,U)
 \ \exp\left(-S[\overline{\psi},\psi,U]\right)},
\eeq
where the action in imaginary time has the form
\beq
  S[\overline{\psi},\psi,U] = \overline{\psi}\ K[U]\ \psi + S_G[U],
\eeq
and where $K[U]$ is the fermion Dirac matrix and $S_G[U]$ is the gluon action.
The integrals over the Grassmann-valued quark/antiquark fields can be done exactly,
leaving expressions of the form
  \beq
  C_{ij}(t)= \frac{ \int {\cal D}U\ \det K[U]\ 
  \left( K^{-1}[U]\cdots K^{-1}[U] +\dots\right)\ \ \exp\left(-S_G[U]\right)}{  
  \int {\cal D}U\ \det K[U]
 \ \exp\left(-S_G[U]\right)}.
  \eeq
For the remaining integrations over the gluon fields, the Monte Carlo method is used.
This requires formulating QCD on a space-time lattice (usually hypercubic),
with quark fields residing on the sites and the gluon field residing on the links 
between lattice sites.  The lattice QCD action is formulated in such a way so as to
maintain local gauge invariance\cite{Wilson:1974sk}.  A Markov chain is used to 
generate a sequence of gauge-field
configurations $ U_1, U_2,\dots, U_N$ using the Metropolis method\cite{Metropolis:1953am}
with a complicated global updating proposal, such as RHMC\cite{Clark:2006fx}, 
which proposes a new gauge field by selecting conjugate momenta randomly chosen
with Gaussian distributions and evolving the fields in a fictitious time
variable using Hamilton equations.  The proposed field differs globally from 
the current field, but the value of its action is very near that of the
current field, ensuring a high acceptance rate.
The fermionic determinants $\det K$ are usually estimated using a
multivariate Gaussian integral over so-called pseudo-fermion fields.
The correlators are then estimated using the ensemble of gauge configurations
generated by the above procedure.  Systematic errors include discretization and
finite volume effects.  To speed up computations, unphysically large quark masses
are often used.

\begin{figure}
 \begin{center}
 \includegraphics[width=3.5in]{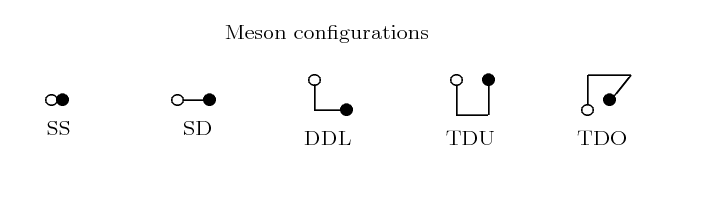}\\[-20pt]
 \includegraphics[width=4.0in]{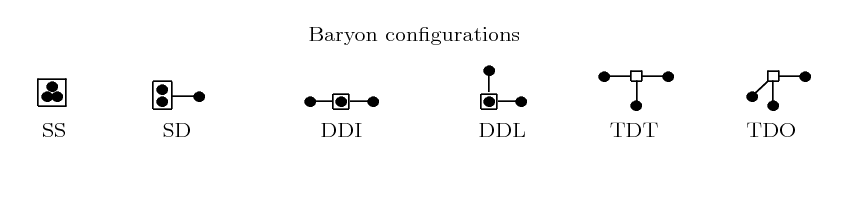}
 \end{center}
\vspace*{-8mm}
\caption{The spatial arrangements of the quark-antiquark meson operators (top)
and the three-quark baryon operators (bottom) that we use. Smeared quarks fields
are shown as solid circles, each hollow 
circle indicates a smeared antiquark field, the solid line segments indicate 
covariant displacements, and each hollow box indicates the location of a 
Levi-Civita color coupling. 
\label{fig:shopers}}
\end{figure}

An efficient way to construct single-hadron operators is to assemble them using
covariantly-displaced smeared quark fields as building blocks.  
Stout link smearing\cite{Morningstar:2003gk} 
is used for the gauge field links $\widetilde{U}_j(x)$,
and Laplacian-Heaviside (LapH)\cite{HadronSpectrum:2009krc,Morningstar:2011ka}
smearing is used for the quark fields
   \beq
    \widetilde{\psi}_{a\alpha}(x) =
      {\cal S}_{ab}(x,y)\ \psi_{b\alpha}(y),
     \qquad {\cal S} = 
     \Theta\left(\sigma_s^2+\widetilde{\Delta}\right),
   \eeq
where $\widetilde{\Delta}$ denotes a 3-dimensional gauge-covariant Laplacian 
defined in terms of the stout links $\widetilde{U}$ and $\sigma_s$ is the
smearing cutoff.  Displaced quark fields are defined by
  \beq
 q^A_{a\alpha j}= D^{(j)}\widetilde{\psi}_{a\alpha}^{(A)},
 \qquad  \overline{q}^A_{a\alpha j}
 = \widetilde{\overline{\psi}}_{a\alpha}^{(A)}
  \gamma_4\, D^{(j)\dagger}
  \eeq
where the displacement $D^{(j)}$ is a product of smeared links
\beq
 D^{(j)}(x,x^\prime) =
 \widetilde{U}_{j_1}(x)\ \widetilde{U}_{j_2}(x\!+\!d_2)
 \ \widetilde{U}_{j_3}(x\!+\!d_3)\dots \widetilde{U}_{j_p}(x\!+\!d_p)
  \delta_{x^\prime,\ x+d_{p+1}}.
\eeq
In the above equations, $a,b$ are color indices, $\alpha$ is a Dirac spin
index, and $j$ labels a displacement path of gauge links in directions $j_1,
j_2,\cdots$.
A variety of displacements can be used to build up the needed orbital and
radial structure, as shown in Fig.~\ref{fig:shopers}.  So-called elemental
quark-antiquark and three-quark operators which create a definite momentum 
$\pvec$ are defined by
 \beqs
 \overline{\Phi}_{\alpha\beta}^{AB}(\pvec,t)&=&
 \textstyle\sum_{\bm{x}} e^{i\pvec\cdot(\xvec+\frac{1}{2}(\bm{d}_\alpha+\bm{d}_\beta))}
   \delta_{ab}\ \overline{q}^B_{b\beta}(\bm{x},t)\ q^A_{a\alpha}(\bm{x},t),
 \\
  \overline{\Phi}_{\alpha\beta\gamma}^{ABC}(\pvec,t)&=& 
 \textstyle\sum_{\bm{x}} e^{i\pvec\cdot\xvec}\varepsilon_{abc}
\ \overline{q}^C_{c\gamma}(\bm{x},t)
\ \overline{q}^B_{b\beta}(\bm{x},t)
\ \overline{q}^A_{a\alpha}(\bm{x},t).
\eeqs
In these operators, $A,B,C$ are quark flavor indices, $a,b,c$ are color indices,
$\bm{d}_\alpha, \bm{d}_\beta$ are the spatial displacements of the $q, \overline{q}$
fields, respectively, from site $\bm{x}$,
and $\alpha,\beta,\gamma$ denote compound indices incorporating both
a Dirac spin index and a displacement path.
Group theoretical projections onto the irreducible representations (irreps) 
of the lattice symmetry group are then employed to create the final
single meson and single baryon operators:
\beq
  \overline{M}_{l}(t)= c^{(l)\ast}_{
 \alpha\beta}\ \overline{\Phi}^{AB}_{\alpha\beta}(t)\qquad\qquad
  \overline{B}_{l}(t)= c^{(l)\ast}_{
 \alpha\beta\gamma}\ \overline{\Phi}^{ABC}_{\alpha\beta\gamma}(t).
\eeq
In the above, $\alpha,\beta,\gamma$ again indicate compound indices incorporating both
a Dirac spin index and a displacement path, and $l$ is the final index
which labels the hadron operator.

Implementing two- and three-hadron operators as appropriate superpositions 
of products of the single-hadron operators of definite momenta is straightforward:
    \beq
    c^{I_{3a}I_{3b}}_{\pvec_a\lambda_a;\ \pvec_b\lambda_b}
     \ B^{I_aI_{3a}S_a}_{\pvec_a\Lambda_a\lambda_a i_a}
     \  B^{I_bI_{3b}S_b}_{\pvec_b\Lambda_b\lambda_b i_b}
    \eeq
for fixed total momentum $\pvec=\pvec_a+\pvec_b$ and
fixed $\Lambda_a, i_a, \Lambda_b, i_b$.  Group theory projections onto 
the little group of $\pvec$ and isospin irreps are then carried out.
It is very important to specify all phases of the single-hadron operators 
for all momenta, and this is generally done by selecting a reference
momentum direction $\pvec_{\rm ref}$, then for each momentum $\pvec$, 
selecting one reference rotation $R_{\rm ref}^{\pvec}$ that 
transforms $\pvec_{\rm ref}$ into $\pvec$.  This method creates large
numbers of multi-hadron operators very efficiently.

Once the temporal correlations are obtained, their spectral representations
can be used to extract the stationary-state energies:
   \beq
   C_{ij}(t) = \sum_n Z_i^{(n)} Z_j^{(n)\ast}\ e^{-E_n t},
   \qquad\quad Z_j^{(n)}=  \me{0}{O_j}{n},
   \label{eq:spectro_decomp}
   \eeq
which neglects small temporal wrap-around contributions, where the 
energies $E_n$ are discrete in finite volume. 
Given the large number of complex parameters in Eq.~(\ref{eq:spectro_decomp}),
it is not feasible to fit the entire correlation matrix using
Eq.~(\ref{eq:spectro_decomp}).  Some method of rotating or diagonalizing
the matrix is needed to make the extraction of the energies and
overlap factors by least-squares fitting practical.  The use of 
variational techniques to assist with this dates back to 
Refs.~\cite{Fox:1981xz,Ishikawa:1982tb,Michael:1982gb,corrmatt2}.
The way we proceed is to define a new
correlation matrix $\widetilde{C}(t)$ using a single rotation
\beq
 \widetilde{C}(t) = U^\dagger\ C(\tau_0)^{-1/2}\ C(t)\ C(\tau_0)^{-1/2}\ U,
 \label{eq:Crotate}
\eeq
where the columns of $U$ are the eigenvectors of
   $C(\tau_0)^{-1/2}\,C(\tau_D)\,C(\tau_0)^{-1/2}$.
One then chooses $\tau_0$ and $\tau_D$ large enough so $\widetilde{C}(t)$ 
remains diagonal for $t>\tau_D$ within statistical errors.
Two-exponential fits to the diagonal rotated correlators
$\widetilde{C}_{\alpha\alpha}(t)$ then yield the energies $E_\alpha$ and 
overlaps $Z_j^{(n)}$. Energy shifts from non-interacting values can also
be obtained from single exponential fits to a suitable ratio of 
correlators, but such fits must be cautiously done in combination with
fits to correlators that are not ratios.

\begin{figure}[t]
\begin{center}
\includegraphics[width=1.48in]{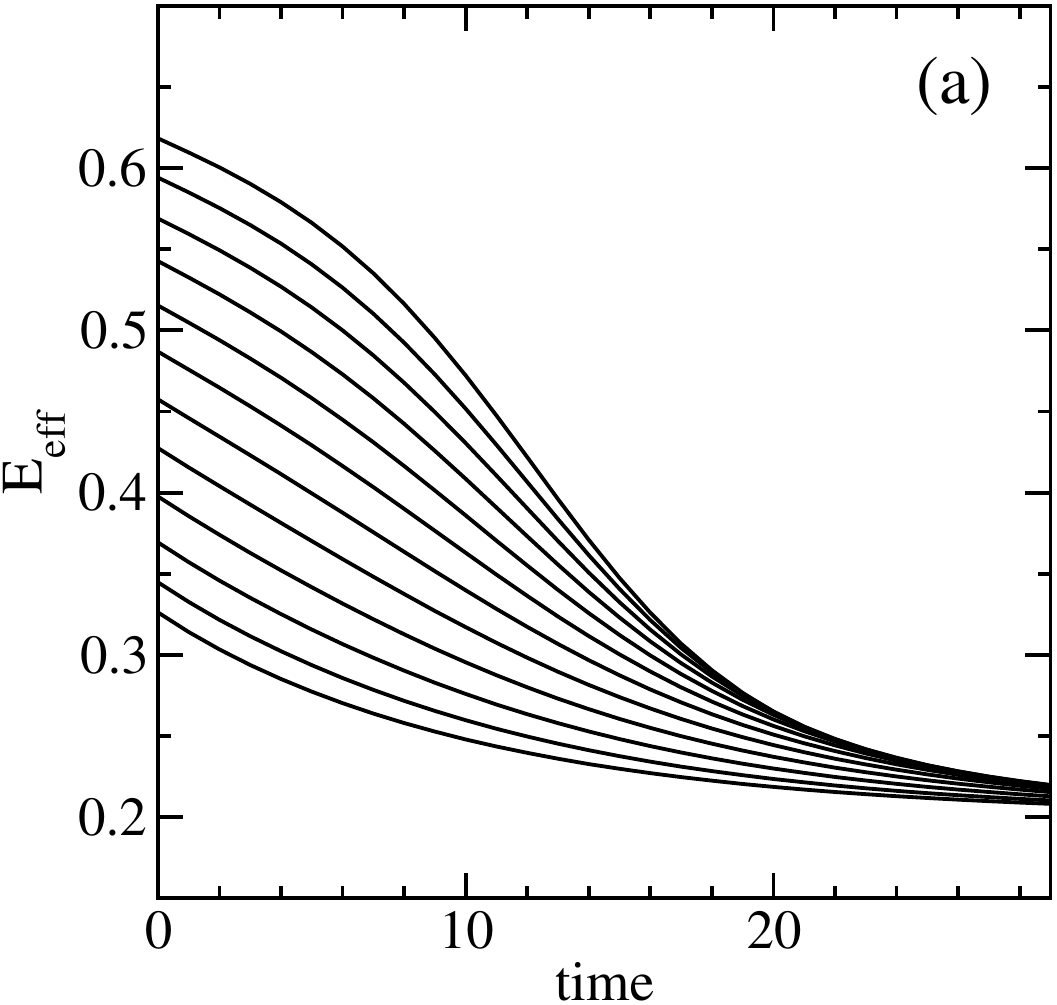}
\includegraphics[width=1.5in]{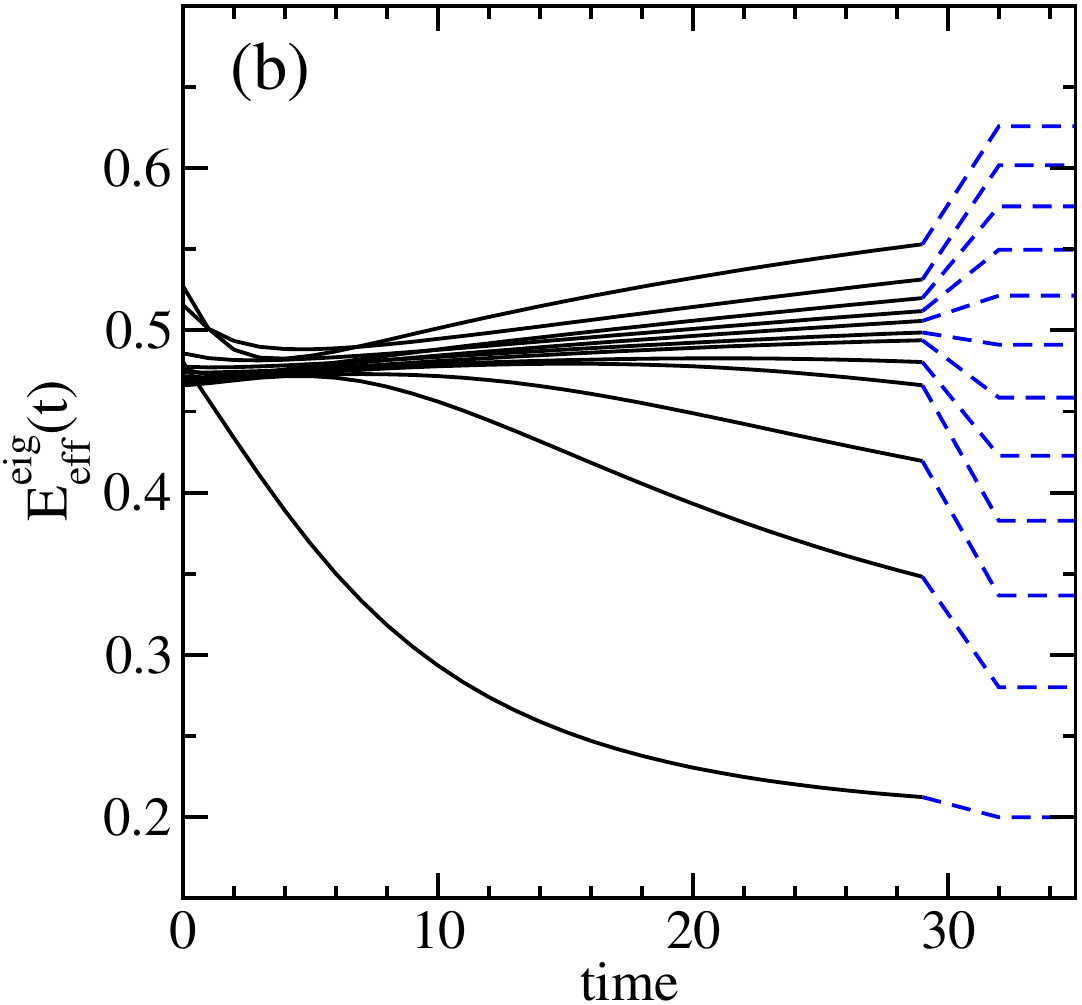}
\includegraphics[width=1.5in]{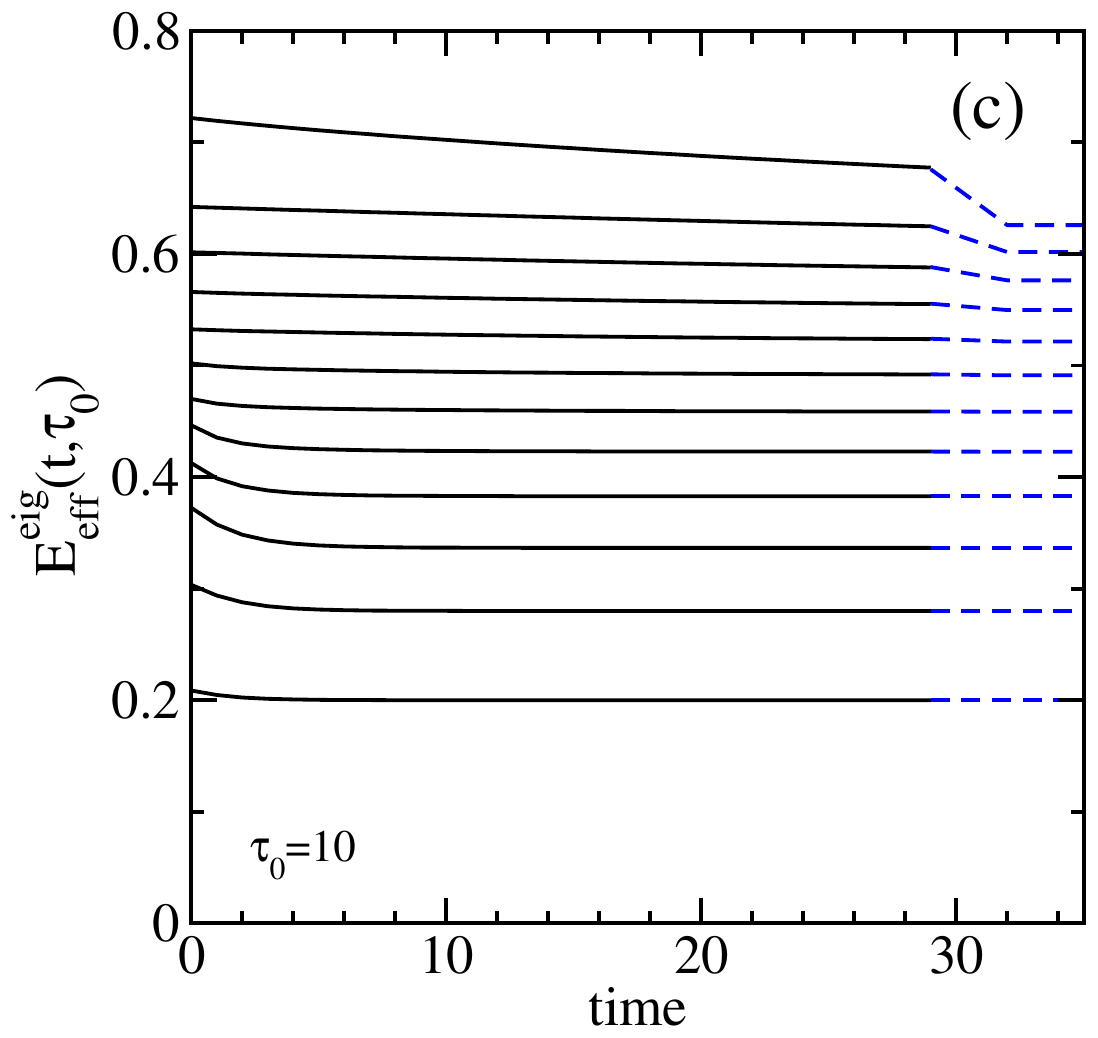}
\end{center}
\caption{(a) Effective energies associated with the diagonal elements
of the original raw correlator matrix $C(t)$ of the toy model, whose energies
are defined in Eq.~(\ref{eq:toy1}) and the overlaps in Eq.~(\ref{eq:toy2}).
(b) Effective energies associated with the eigenvalues of the original
correlator matrix $C(t)$ with eigenvector pinning used to label or order
the eigenvalues. (c) Effective energies associated
with the eigenvalues of $C(\tau_0)^{-1/2}C(t)C(\tau_0)^{-1/2}$ with 
eigenvector pinning, for $\tau_0=10$.
\label{fig:toy1}}
\end{figure}

The usefulness of Eq.~(\ref{eq:Crotate}) can be demonstated using
a simple toy model example.  In this example, there are
$N_e=200$ eigenstates $\ket{n}$ for $n=0,1,\dots,N_e-1$,
and the energies are taken to be
\beq
E_0=0.20,\qquad E_n=E_{n-1}+\frac{0.08}{\sqrt{n}}, \qquad n=1,2,\dots,N_e-1.
\label{eq:toy1}
\eeq
This example studies an $N\times N$ correlator matrix, where $N=12$.  A set
of operators is selected such that each operator
is expected to predominantly create a different eigenstate, so the
overlaps of the $j$-th operator onto the eigenstates is taken to be
\beq
Z_j^{(n)}=\frac{(-1)^{j+n}}{1+0.05(j-n)^2}.
\label{eq:toy2}
\eeq 
The so-called ``effective energies'' associated with the diagonal elements
of the original raw correlator matrix of this toy model are shown
in Fig.~\ref{fig:toy1}(a). Each effective energy is defined by 
\beq
E_{\rm eff}^{(n)}(t)=\ln\left(\frac{C_{nn}(t)}{C_{nn}(t+1)}\right).
\eeq
Each of these effective energies eventually tends down to $E_0=0.20$
since the state created by each operator has some overlap with
the lowest-lying eigenstate $\ket{0}$.  As the operator index increases,
the couplings to higher-lying eigenstates get larger, so one sees
that the effective energies associated with those operators have
higher values at smaller times.

The effective energies associated with the eigenvalues $\lambda_n(t)$ of $C(t)$ are
shown in Fig.~\ref{fig:toy1}(b).  Each of these effective energies is defined by 
\beq
E_{\rm eff}^{{\rm eig}(n)}(t)=\ln\left(\frac{\lambda_{n}(t)}{\lambda_{n}(t+1)}\right).
\eeq
To order the eigenvalues for different $t$ values, eigenvector
``pinning" has been used.  For one reference value of time $t_{\rm ref}$, 
a particular order of the eigenvalues has been chosen, and the eigenvectors 
associated with these eigenvalues are used as references.  For a different 
value of time $t\neq t_{\rm ref}$, the inner products of the eigenvectors at 
$t$ are taken with those at the reference time $t_{\rm ref}$, and the
eigenvalue associated with the eigenvector at time $t$ with maximal overlap 
onto reference eigenvector $n$ is identify as $\lambda_n(t)$.
The dashed blue lines indicate
the lowest 12 exact energies.  These effective energies will eventually
tend toward the horizontal dashed blue lines, but certainly by $t=30$,
most of these effective energies still significantly disagree with
their expected large-$t$ values.  Also, many of the curves cross each
other, and the curves are very close to one another for small and
moderate $t$ values.  Clearly, trying to extract the spectrum of energies
from the eigenvalues of $C(t)$ will be a difficult task.

The effective energies associated with the eigenvalues $\lambda_n(t,\tau_0)$
of the modified correlator matrix
$C(\tau_0)^{-1/2}C(t)C(\tau_0)^{-1/2}$ are shown in Fig.~\ref{fig:toy1}(c)
for $\tau_0=10$. Each of these effective energies is defined by 
\beq
E_{\rm eff}^{{\rm eig}(n)}(t,\tau_0)
 =\ln\left(\frac{\lambda_{n}(t,\tau_0)}{\lambda_{n}(t+1,\tau_0)}\right).
\eeq
To order the eigenvalues, eigenvector ``pinning" has again been used.
The dashed blue lines indicate the lowest 12 exact energies. 
These effective energies look dramatically different from those in
Fig.~\ref{fig:toy1}(b).  The approach of these effective energies to
the dashed blue lines is dramatically faster, and no crossings of the
curves are observed.  By time $t=30$, the lower-lying levels already
agree with their expected large-$t$ values, and only a few of the
levels near $N$ still exhibit a significant difference from its
limiting blue line.  These results clearly show the great superiority
of using the eigenvalues of $C(\tau_0)^{-1/2}C(t)C(\tau_0)^{-1/2}$ 
instead of $C(t)$.  If one varies $\tau_0$, one finds that the results 
are insensitive to the value of $\tau_0$.  
One also sees that if $N$ levels are desired, then a correlator matrix
larger than $N\times N$ should be used, such as 
$\frac{3}{2}N\times \frac{3}{2}N$.  

The relationship of the finite-volume energies obtained in lattice QCD and
the infinite-volume scattering $S$-matrix was first studied in detail in  
Refs.~\cite{Luscher:1990ux,Luscher:1991cf}, but lattice QCD computations
at that time could not determine the energies of multi-hadron states reliably and
accurately enough to take advantage of this relationship.  As lattice QCD
methodology improved for multi-hadron operators, the relationship between
scattering amplitudes and finite-volume stationary-state energies was
revisited in Refs.~\cite{Rummukainen:1995vs,Kim:2005gf}, limiting
attention to a single channel of identical spinless particles.  Later works
generalized their results to treat multi-channels with different particle
masses and nonzero spins\cite{Briceno:2014oea}. Our procedure for evaluating
scattering phase shifts from lattice QCD energies is presented in
Ref.~\cite{Morningstar:2017spu} and is summarized below.

Instead of the unitary $S$-matrix, the real and symmetric 
$K$-matrix\cite{Wigner:1946zz,Wigner:1947zz}, defined using the $S$-matrix by
\beq
   S = (1+iK)(1-iK)^{-1} = (1-iK)^{-1}(1+iK),
\eeq
is usually used in the quantization condition that relates the scattering
amplitudes and the finite-volume energies since it is easier to parametrize 
a real symmetric matrix than
a unitary matrix. Rotational invariance implies that
\beq 
  \langle J'm_{J'}L^\prime S^\prime a'\vert\ K
\ \vert Jm_JLS  a\rangle = \delta_{J'J}\delta_{m_{J'}m_J}
 \ K^{(J)}_{L'S'a';\ LS a}(\Ecm).
\eeq
We use an orthonormal basis of states, each labelled by $\ket{Jm_JLS a}$, where $J$ is
the total angular momentum of the two particles in the center-of-momentum frame, $m_J$ is 
the projection of the total angular momentum onto the $z$-axis, $L$ is the orbital angular 
momentum of the two particles in the center-of-momentum frame (not to be confused with the
lattice length $L_b$ here), and $S$ in the basis vector is 
the total spin of the two particles (not the scattering matrix).  
The multichannel generalization\cite{Ross:1961jlg,deSwart:1962,Burke:2011}
of the effective range expansion (ERE) is
\beq
 K^{{-1}(J)}_{L'S'a';\ LSa}(\Ecm)=q_{{\rm cm},a'}^{-L'-\frac{1}{2}}
 \ {\widetilde{K}}^{{-1}(J)}_{L'S'a';\ LSa}(\Ecm)
  \ q_{{\rm cm},a}^{-L-\frac{1}{2}},
\label{eq:Keffrange}
\eeq
where ${\widetilde{K}}^{{-1}(J)}_{L'S'a';\ LSa}(\Ecm)$ is a real, symmetric, and often
analytic function of the center-of-momentum energy $\Ecm$. 
For a given total momentum $\Pvec=(2\pi/L_b)\dvec$ in a spatial $L_b^3$ volume
with periodic boundary conditions, where $\dvec$ is a vector of integers, 
we determine the total lab-frame energy $E$ for a
two-particle interacting state in our lattice QCD simulations.  If the 
masses of the two particles in decay channel $a$ are $m_{1a}$ and $m_{2a}$, we boost to the 
center-of-mass frame and define
\begin{eqnarray}
   \Ecm &=& \sqrt{E^2-\Pvec^2},\qquad
   \gamma = \frac{E}{\Ecm},\\
   \qqa &=& \frac{1}{4} \Ecm^2
   - \frac{1}{2}(m_{1a}^2+m_{2a}^2) + \frac{(m_{1a}^2-m_{2a}^2)^2}{4\Ecm^2},\nonumber\\
   u_a^2&=& \frac{L^2\qqa}{(2\pi)^2},\qquad
 \svec_a = \left(1+\frac{(m_{1a}^2-m_{2a}^2)}{\Ecm^2}\right)\dvec.
\end{eqnarray}
The total lab-frame energy $E$ is related to the scattering $K$-matrix
through the quantization condition:
\beq
\det(1-B^{(\Pvec)}\widetilde{K})=\det(1-\widetilde{K}B^{(\Pvec)})=0,\qquad
  \det(\widetilde{K}^{-1}-B^{(\Pvec)})=0.
\label{eq:quant}
\eeq
The \textit{box matrix} is given by
\beqs
 && \me{J'm_{J'}L'S'a'}{B^{(\Pvec)}}{Jm_JLS a} =
-i\delta_{a'a}\delta_{S'S} \ q_{{\rm cm},a}^{L'+L+1}\ W_{L'm_{L'};\ Lm_L}^{(\Pvec a)}  \nn\\
&&\qquad\qquad \times\langle J'm_{J'}\vert L'm_{L'},Sm_{S}\rangle
\langle Lm_L,Sm_S\vert Jm_J\rangle.
\label{eq:Bmatdef}
\eeqs
This box matrix $B^{(\Pvec)}$ is Hermitian for $q_{{\rm cm},a}^2$ real,
and the determinants in Eq.~(\ref{eq:quant}) are real.  
The $\langle j_1m_1 j_2m_2\vert JM\rangle$ are the familiar Clebsch-Gordan coefficients,
and the $W^{(\Pvec a)}$ matrix elements are given by
\beqs
-iW^{(\Pvec a)}_{L'm_{L'};\ Lm_L} 
&=& \sum_{l=\vert L'-L\vert}^{L'+L}\sum_{m=-l}^l
   \frac{ {\cal Z}_{lm}(\svec_a,\gamma,u_a^2) }{\pi^{3/2}\gamma u_a^{l+1}}
\sqrt\frac{(2{L'}+1)(2l+1)}{(2L+1)}\nonumber\\
&& \qquad\qquad\times \langle {L'} 0,l 0\vert L 0\rangle 
\langle {L'} m_{L'},  l m\vert  L m_L\rangle.
\eeqs
The Rummukainen-Gottlieb-L\"uscher (RGL) shifted zeta functions are evaluated
using
\beqs
   &&{\cal Z}_{lm}(\svec,\gamma,u^2)=\sum_{\nvec\in \bm{Z}^3}
  \frac{{\cal Y}_{lm}(\zvec)}{(\zvec^2-u^2)}e^{-\Lambda(\zvec^2-u^2)}
 +\delta_{l0}\frac{\gamma\pi}{\sqrt{\Lambda}} F_0(\Lambda u^2)
\nn\\
 &&\qquad +\frac{i^l\gamma}{\Lambda^{l+1/2}} \int_0^1\!\!dt 
\left(\frac{\pi}{t}\right)^{l+3/2}\! e^{\Lambda t u^2}
\sum_{\nvec\in \bm{Z}^3\atop \nvec\neq 0}
e^{\pi i \nvec\cdot\svec}{\cal Y}_{lm}(\wvec)
\  e^{-\pi^2\wvec^2/(t\Lambda)},
\label{eq:zaccfinal}
\eeqs
where $\zvec= \nvec -\gamma^{-1} \bigl[\textstyle\frac{1}{2}
+(\gamma-1)s^{-2}\nvec\cdot\svec \bigl]\svec$ and
$\wvec=\nvec - (1  - \gamma) s^{-2}
 \svec\cdot\nvec\svec$, the spherical harmonic polynomials are given by
${\cal Y}_{lm}(\xvec)=\vert \xvec\vert^l\ Y_{lm}(\widehat{\xvec})$,
and 
\begin{equation}
  F_0(x) =  -1+\frac{1}{2}
\int_0^1\!\! dt\ \frac{e^{tx}-1 }{t^{ 3/2}}.
\end{equation}
We choose $\Lambda\approx 1$ which allows sufficient
convergence speed of the summations. 

To use the determinants in Eq.~(\ref{eq:quant}) in practice, we transform to
a block-diagonal basis and truncate in orbital angular momentum.  Matrices corresponding
to symmetry operations in the little group of $\Pvec$ commute with the box matrix,
leading to block-diagonal basis states
\beq
  \ket{\Lambda\lambda n JLS a}= \sum_{m_J} c^{J(-1)^L;\,\Lambda\lambda n}_{m_J} 
 \ket{Jm_JLS a},
\eeq
where $\Lambda$ denotes an irrep of the little group, $\lambda$ labels the row of the
irrep, and $n$ is an occurrence index.  The transformation coefficients depend on $J$ and 
$(-1)^L$, but not on $S,a$.  In this block-diagonal basis, the box matrix and the
$\widetilde{K}$ matrix for $(-1)^{L+L'}=1$ have the forms
\beqs
 \me{\Lambda'\lambda' n'J'L'S' a'}{B^{(\Pvec)}}{\Lambda\lambda nJLS a}
 &=& \delta_{\Lambda'\Lambda}\delta_{\lambda'\lambda}\delta_{S'S}
 \delta_{a'a}\ B^{(\Pvec\Lambda_B Sa)}_{J'L'n';\ JLn}(\Ecm), \nonumber\\
 \me{\Lambda'\lambda' n'J'L'S' a'}{\widetilde{K}}{\Lambda\lambda nJLS a}
&=& \delta_{\Lambda'\Lambda}\delta_{\lambda'\lambda}\delta_{n'n} \delta_{J'J}
\ \widetilde{K}^{(J)}_{L'S'a';\ LS a}(\Ecm). \nonumber
\eeqs
To eliminate any dependence on the truncation in orbital angular momentum,
one can keep increasing the maximum retained $L$ until the results converge.

The determinant in Eq.~(\ref{eq:quant}) is not a good quantity in which to 
search for zeros since it can vary rapidly and can become very large in magnitude.
Furthermore, the box matrix is divergent at all of the non-interacting
two-particle energies.  When computing bootstrap errors, some of resamplings can
cross these singularities, dramatically magnifying statistical errors.
More importantly, when interactions are fairly weak in
a scattering process, the required solutions of the quantization condition can occur
uncomfortably close to these noninteracting energies where the divergences occur,
making root finding to locate the zeros difficult.

The problem of the determinant becoming very large in magnitude is straightforward 
to deal with. Expressing the quantization condition 
in terms of the determinant is just a convenient way of saying that one eigenvalue 
becomes zero.  One obvious way to proceed is to simply search for the zeros
in each of the eigenvalues.  Alternatively, consider the following
function of matrix $A$ which appropriately rescales the determinant using a scalar 
parameter $\mu\neq 0$:
\beq
   \Omega(\mu,A)\equiv \frac{\det(A)}{\det[(\mu^2+AA^\dagger)^{1/2}]}.
\eeq
To evaluate this function, one first finds the eigenvalues $\lambda_\alpha(A)$
of $A$, then
\beq
  \Omega(\mu,A) = \prod_k\frac{\lambda_k(A)}{(\mu^2+\vert\lambda_k(A)\vert^2)^{1/2}}.
\eeq
Clearly, when one of the eigenvalues is zero, this function is also zero. 
For eigenvalues which are much smaller than $\vert\mu\vert$, the
associated term in the product tends towards the eigenvalue itself,
divided by $\vert\mu\vert$.  However, the key feature of this function is
that for eigenvalues which are much larger than $\vert\mu\vert$, the
associated term in the product goes to $e^{i\theta}$ for real $\theta$.  
This function replaces the large eigenvalues with unimodular quantities
so that the product should never overflow.  This is a much 
better behaved function which still reproduces the quantization condition.
As far as the quantization condition cares, the choice of $\mu$ is
irrelevant. 

The use of a Cayley transform solves the problem of the singularities
in the box matrix.  If we introduce the following Cayley transforms,
\beqs
      \Cb &=& (1+i\Bp)(1-i\Bp)^{-1}=(1-i\Bp)^{-1}(1+i\Bp),\\
      \St &=& (1+i\Kt)(1-i\Kt)^{-1}=(1-i\Kt)^{-1} (1+i\Kt),\\
          &=& -(1-i\Ktinv)(1+i\Ktinv)^{-1},
\eeqs
then a tamed quantization condition can be obtained
\beq
    \det( 1+\St\Cb )=0,\qquad \det( \Stinv+\Cb )=0.
    \label{eq:quantSC}
\eeq
This determinant is no longer real, so both the real and imaginary parts
must be combed for zeros.  The $\Omega$ function can still be applied.

Each quantization condition in Eqs.~(\ref{eq:quant}) or (\ref{eq:quantSC})
is a single relation between an energy $E$ determined in finite-volume and
the entire $K$-matrix.  When multiple partial waves or multiple
channels are involved, this relation is clearly not sufficient
to extract all of the $K$-matrix elements at the single energy $E$.  The best way to 
proceed is to approximate the $K$-matrix elements using physically motivated 
functions of the energy $\Ecm$ involving a handful of parameters.  Values
of these parameters can then be estimated by appropriate fits
using a sufficiently large number of different energies. 

\section{\boldmath The $\Delta$ Resonance}

\begin{figure}[t]
\begin{center}
 \includegraphics[width=0.95\textwidth]{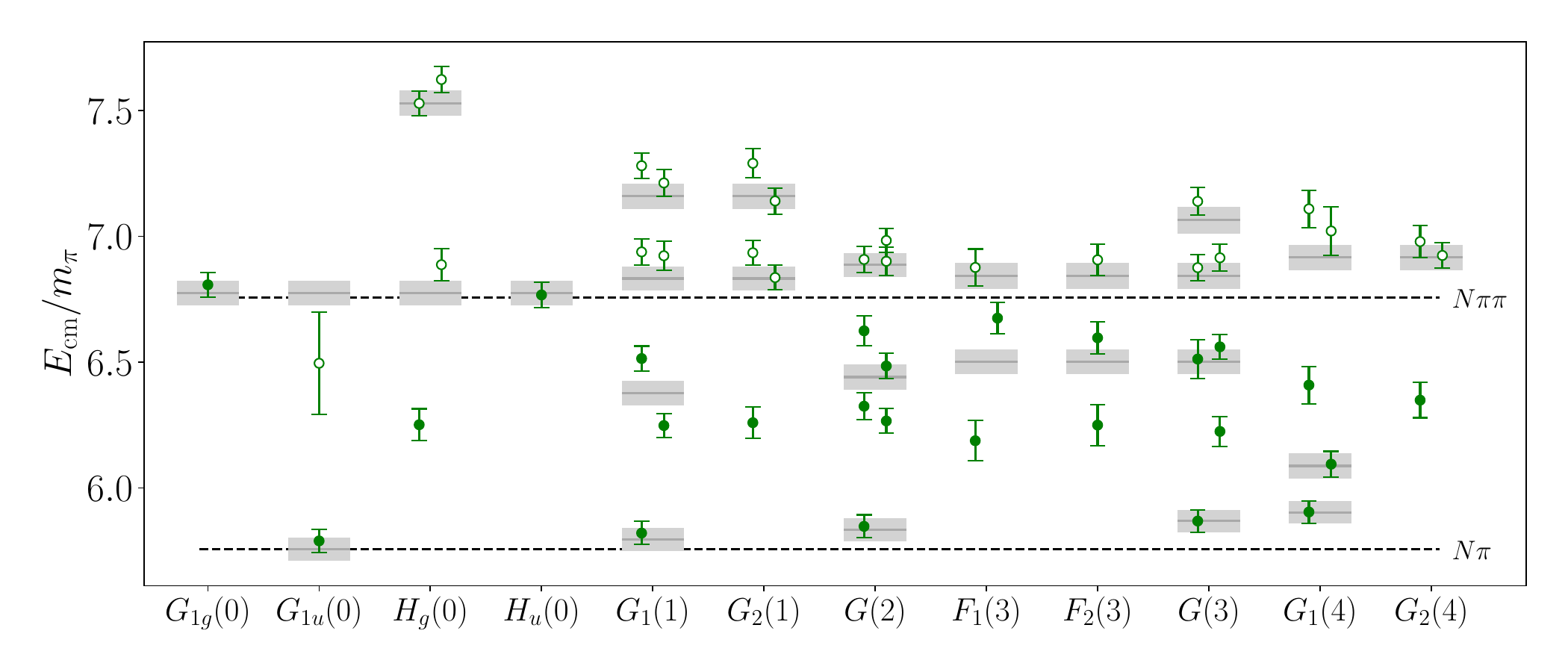}\\
 (a) The $I=3/2$ spectrum.\\
 \includegraphics[width=0.95\textwidth]{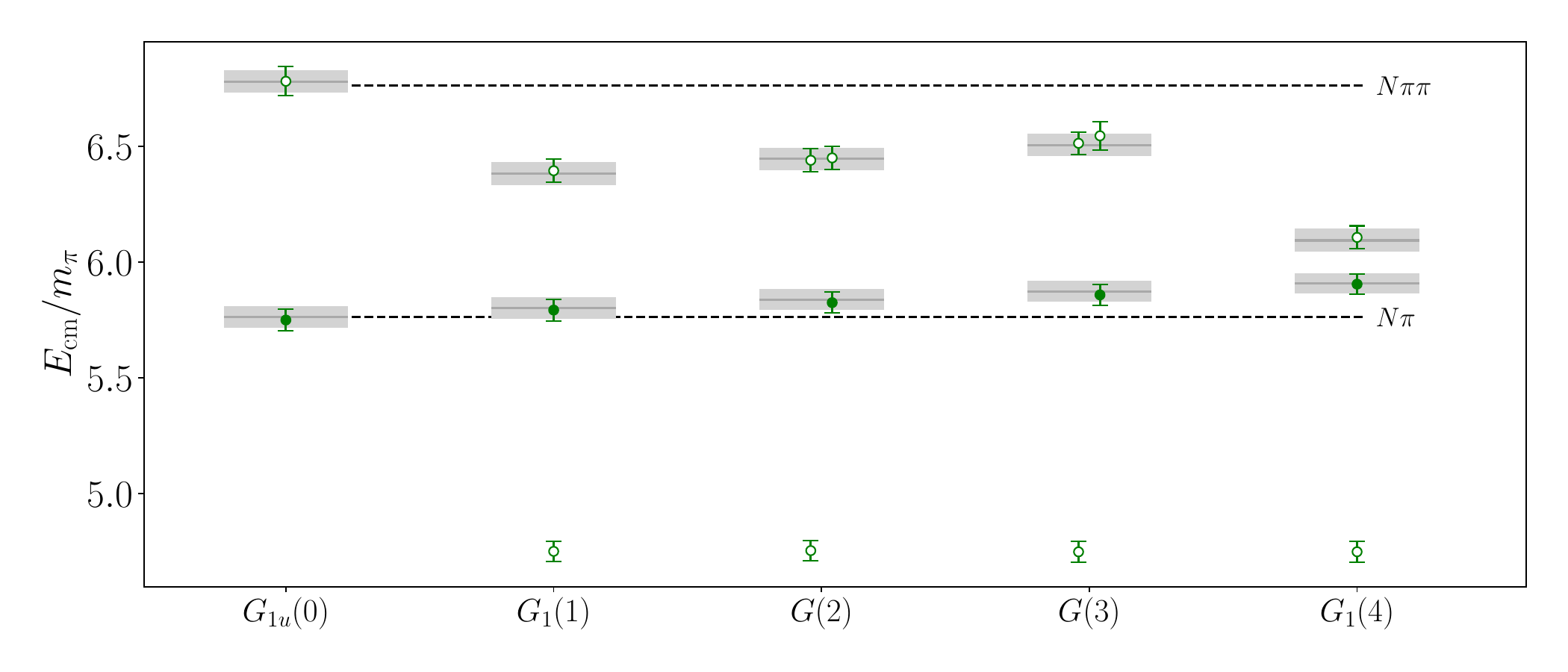}\\
 (b) The $I=1/2$ spectrum.
\end{center}
\caption{The low-lying $I = 3/2$ (top) and $I = 1/2$ (bottom) nucleon-pion spectra in the 
center-of-momentum frame on the D200 ensemble as energies over the pion mass
from Ref.~\cite{Bulava:2022vpq}. Each column corresponds to a 
particular irrep $\Lambda$ of the little group of total momentum 
$\Pvec^2 = (2\pi/L_b)^2 \bm{d}^2$, denoted $\Lambda(\bm{d}^2)$. Dashed lines 
indicate the boundaries of the elastic region. Solid lines and shaded regions
indicate non-interacting $N \pi$ levels and their associated
statistical errors.
\label{fig:deltaspectrum}}
\end{figure}

One of the simplest baryon resonances to study in lattice QCD is the
$\Delta$ resonance, which is an important feature of nucleon-pion scattering.
Our most recent study of $N\pi$ scattering at $m_\pi\sim 200~{\rm MeV}$ was presented
in Ref.~\cite{Bulava:2022vpq}.  Our results were obtained using 2000 configurations
with four source times for the quark propagators of the CLS D200 ensemble, which uses
a $64^3\times 128$ lattice with spacing $a\sim 0.065~\rm{fm}$ and open boundary 
conditions in time.  The quark masses are tuned such that $m_\pi\sim 200~\rm{MeV}$ and 
$m_K\sim 480~{\rm MeV}$. Results for the finite-volume energies
we obtained are shown in Fig.~\ref{fig:deltaspectrum}.  

\begin{figure}[t]
\begin{center}
\includegraphics[width=0.45\linewidth]{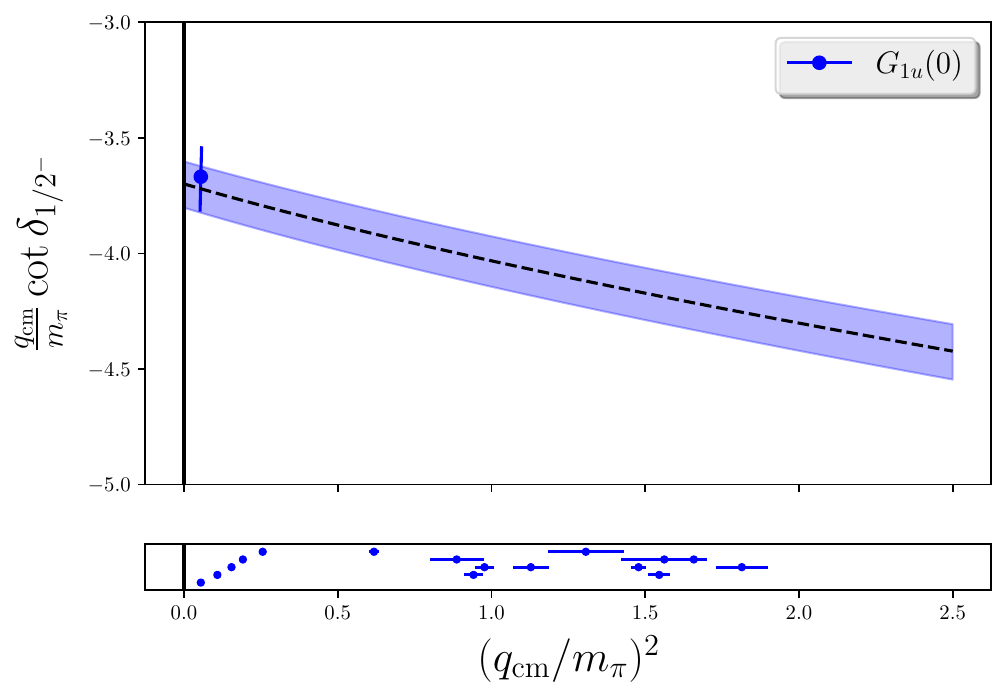}
\includegraphics[width=0.45\linewidth]{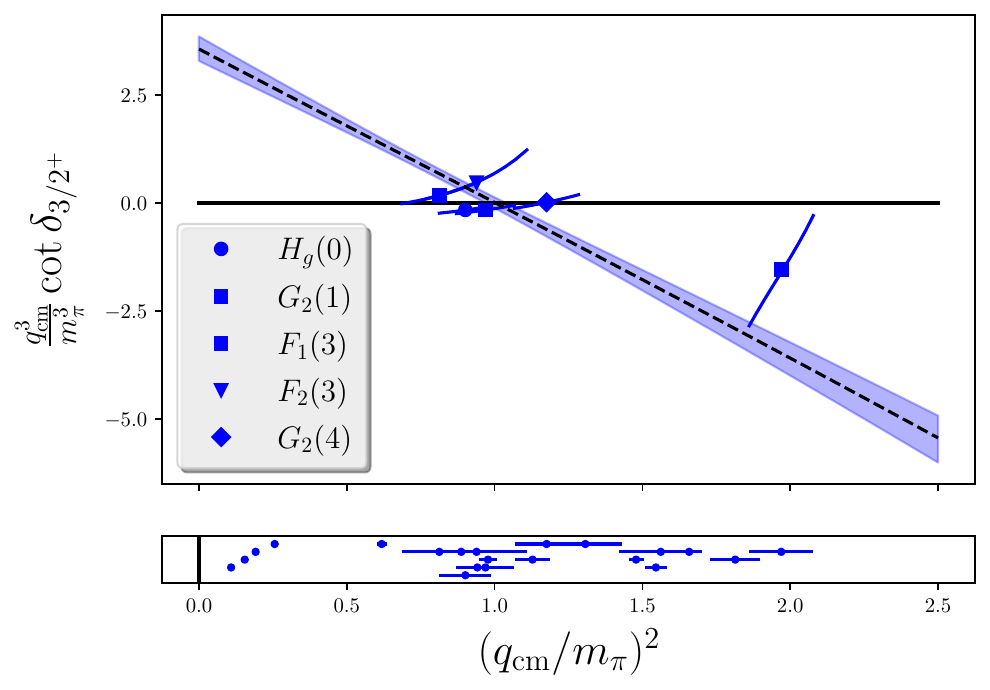}
\includegraphics[width=0.45\linewidth]{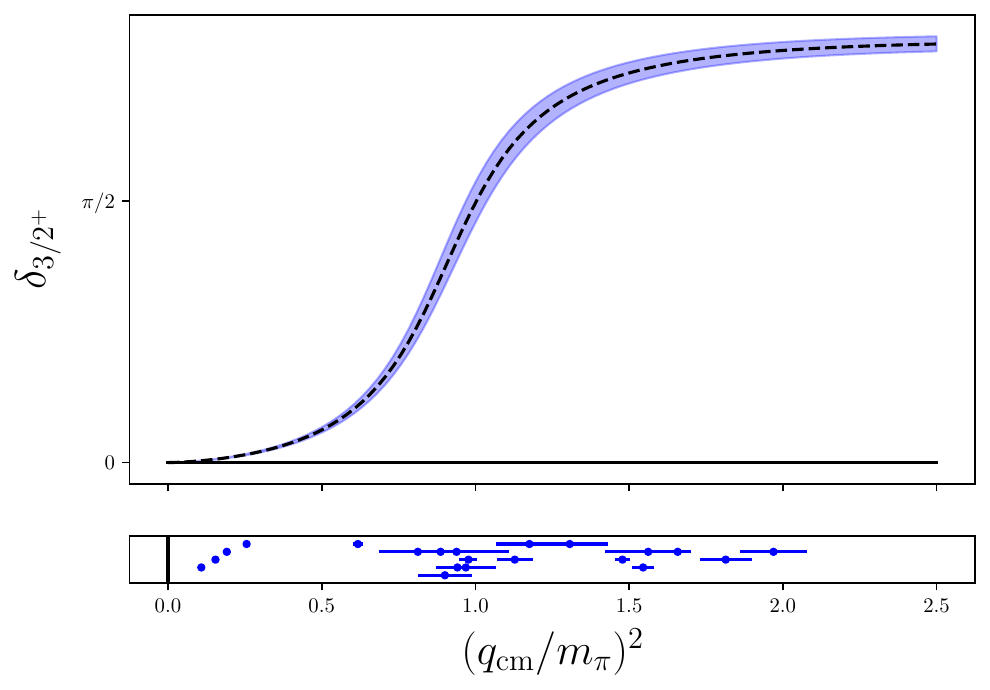}\quad
\raisebox{2mm}{\includegraphics[width=0.45\linewidth]{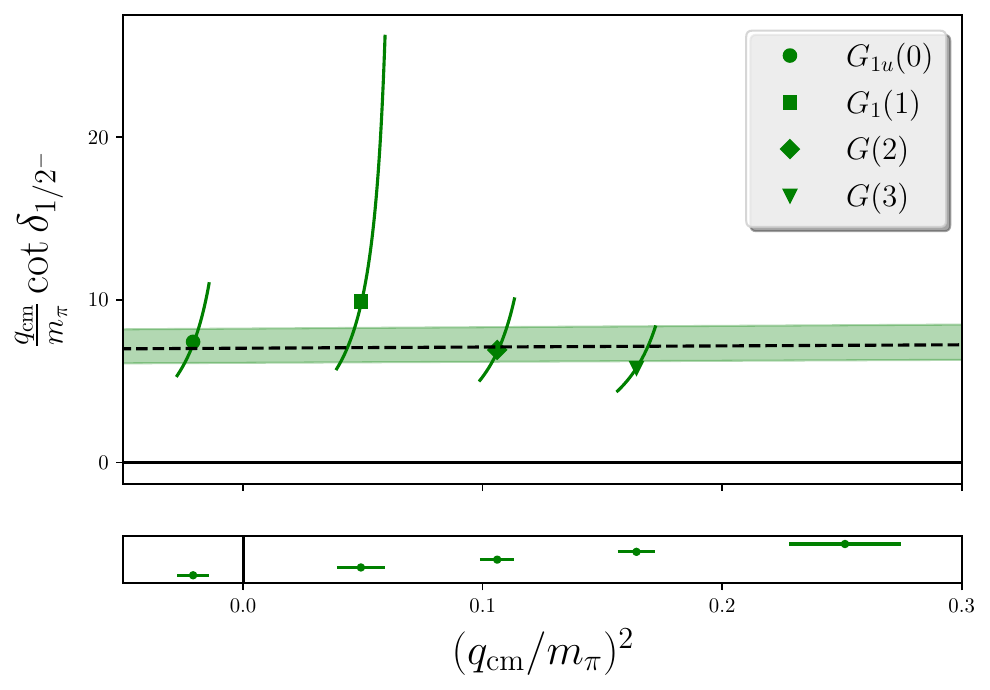}}
\end{center}
\vspace*{-8mm}
\caption{$N\pi$ scattering phase shifts from Ref.~\cite{Bulava:2022vpq}
for $I=\frac{3}{2}$  $s$-wave (top left) and $p$-wave (top right)
in their cotangent form multiplied with threshold momentum factors.  The $p$-wave phase shift itself
 is shown in the bottom left.
 Similarly, $N\pi$ scattering phase shifts for $I=\frac{1}{2}$ $s$-wave (bottom right).
 Lower panels indicate all of the energies used in the fits to obtain the phase shifts in the
 top panels. \label{fig:delta2}}
\end{figure}

The goal of this analysis is a parametrization of the $J^P = 1/2^-$ partial wave for both isospins, 
and the $3/2^+$ wave with $I=3/2$. 
Each partial wave is parametrized using the effective range expansion.  For the $I=3/2$, $J^P=3/2^+$ wave, 
the next-to-leading order is included
\begin{equation}
    \frac{q_{\rm cm}^{3}}{m_\pi^{3}} \cot \delta_{3/2^+}  = \frac{6 \pi \sqrt{s} }{m^3_\pi \gbw^2} 
    ( m_\Delta^2 - s ),
\end{equation}
where $\sqrt{s} = E_{\rm cm}=\sqrt{m_\pi^2 + q_{\rm cm}^2} + \sqrt{m_N^2 + q_{\rm cm}^2}$, and the 
effective range fit parameters are reorganized to form the conventional  Breit-Wigner properties of 
the $\Delta$ resonance, denoted $\gbw^2$ and $m_\Delta$. For the other waves, the leading term 
in the effective range expansion is sufficient
\begin{equation}
    \frac{q_{\rm cm}^{2L+1}}{m_\pi^{2L+1}} \cot \delta^I_{J^P}  = \frac{\sqrt{s}}{m_\pi A^I_{J^P}},
\end{equation}
where the overall $\sqrt{s}$ factors are adopted from standard continuum analysis,
and the single fit parameter $A_{J^P}^I$ is trivially related to the scattering length
\begin{equation}
   m_\pi^{2L+1} a_{J^P}^{I} =  \frac{m_\pi }{m_\pi + m_N} A^I_{J^P} .
\end{equation}

\begin{figure}[p]
\begin{center}
\includegraphics[width=0.95\textwidth]{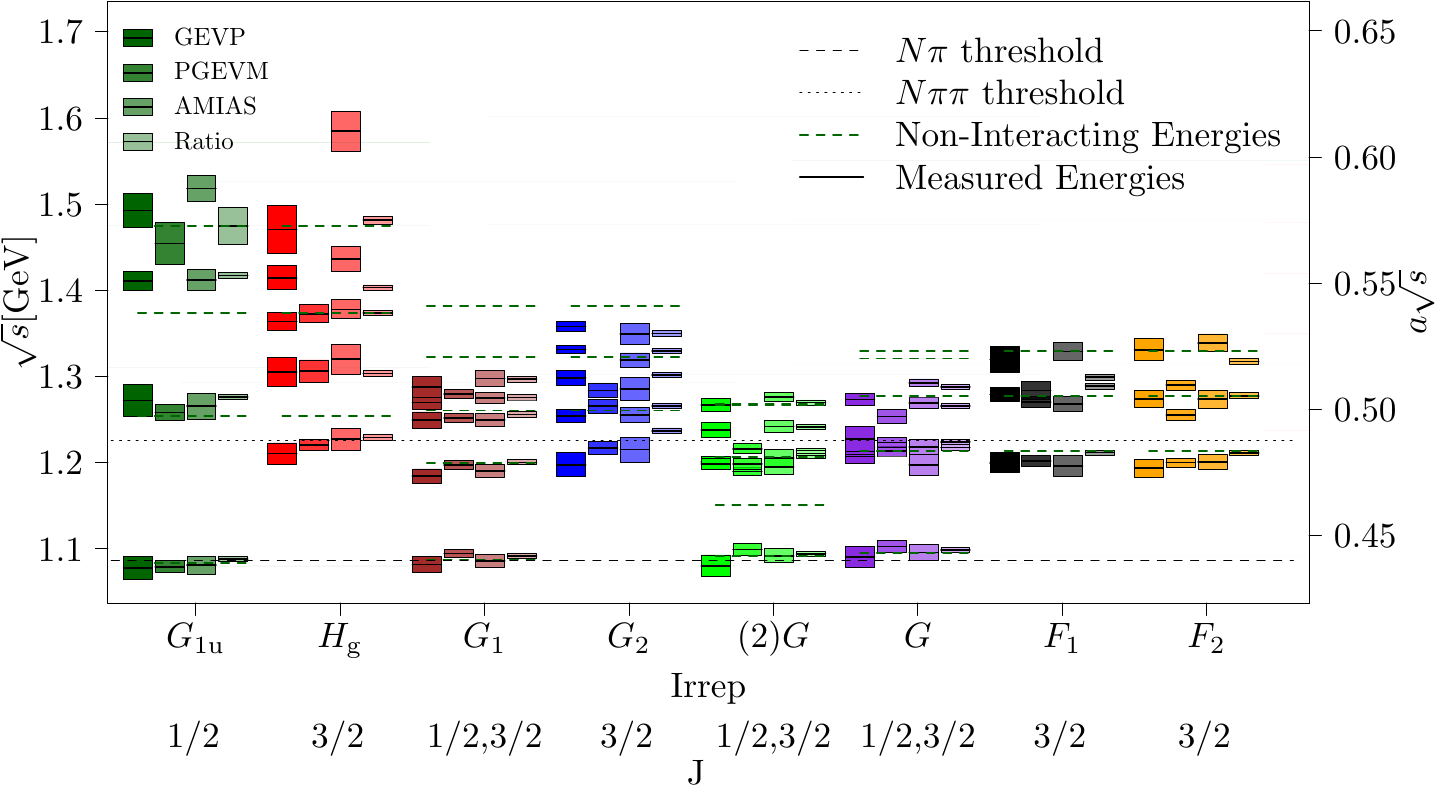}\\[4mm]
\includegraphics[width=0.75\textwidth]{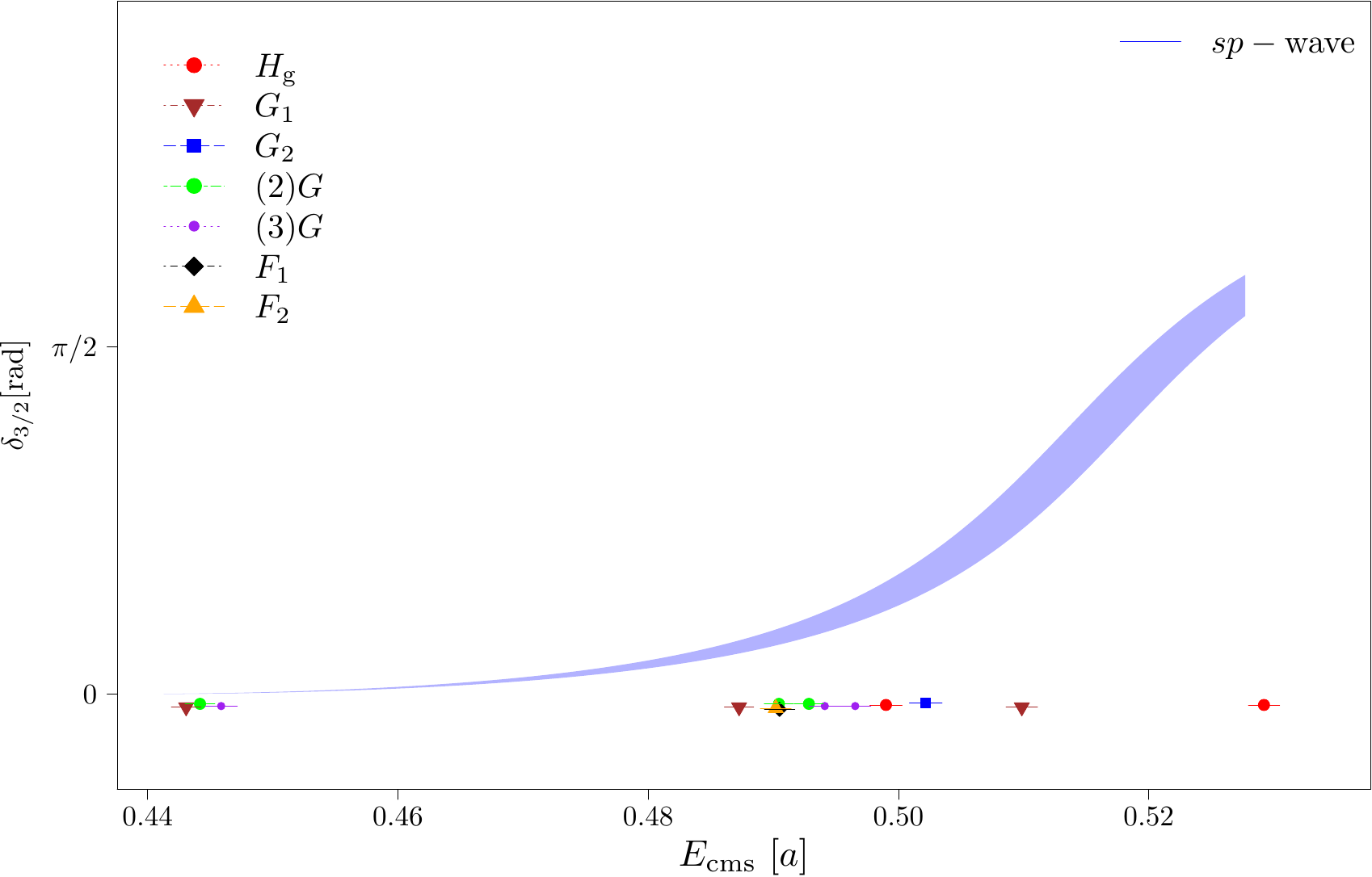}
\end{center}
\caption{(Top) The $\pi N$ interacting two-hadron energy levels obtained in 
Ref.~\cite{Alexandrou:2023elk}.  Box heights indicate estimated uncertainties.
Horizontal dashed/dotted lines show various thresholds, as indicated by the legend.
Noninteracting energies are shown by the green, thicker dashed lines.
(Bottom) The $P$-wave scattering phase-shift as a function of the invariant
mass $E_{\rm cm} = \sqrt{s}$. The error band is determined using jackknife
resampling. The points with horizontal error bars show each fitted
energy level included its jackknife error bar.
\label{fig:physDelta}}
\end{figure}
\clearpage

For the 
$(2J, L) = (3, 1)$ wave, energies in the $H_{g}(0)$, $G_2(1)$, $F_1(3)$, $G_2(4)$
irrep were used.  In each irrep label, the integer in parentheses indicates
$\bm{d}^2$, for total momentum $\Pvec=2\pi\bm{d}/L_b$. The $G_{1u}(0)$ irrep
gives the $(1, 0)$ wave, the irreps used with $s$- and $p$-wave mixing
were $G_1(1)$, $G(2)$, $G_1(4)$. The scattering phase shifts
obtained from the finite-volume energies using the L\"{u}scher quantization
condition are shown in Fig.~\ref{fig:delta2}.  For the $\Delta$ mass and
Breit-Wigner width parameter $g_{\Delta,\rm{BW}}$, as well as the
scattering lengths, the following results were obtained:
\beqs
 \begin{array}{rcl@{\quad}rcl}
 m_\Delta/m_\pi &=& 6.290(18),& g_{\Delta,\rm{BW}}&=&14.7(7), \\
 m_{\pi}a_0^{3/2} &=& -0.2735(81),&  m_{\pi}a_0^{1/2} &=& 0.142(22).
\end{array}
\eeqs
The amplitudes are well-described by the effective range expansion. 

A comparison to chiral perturbation theory is presented in Ref.~\cite{Bulava:2022vpq}.
Not only do our results disagree with leading-order chiral perturbation theory
(LO $\chi$PT), but we find that the magnitude of $m_\pi a_0^{3/2}$ is larger than
that of $m_\pi a_0^{1/2}$, in disagreement with both LO $\chi$PT and
phenomenology.  For more details, see Ref.~\cite{Bulava:2022vpq}.

A more recent study of the $\Delta$ resonance at the physical point (with quark masses 
set to give the physical pion and kaon masses) and lattice spacing $a= 0.08~\rm{fm}$  
was recently presented in Ref.~\cite{Alexandrou:2023elk}.  Their finite-volume
spectrum and scattering phase shift are shown in Fig.~\ref{fig:physDelta}.
Low three-particle thresholds were a problem in
this study.  The $\Delta$ resonance mass and width were found to be
\begin{eqnarray*}
 M_R &=& 1269\,(39)_{\rm{Stat.}}(45)_{\rm{Total}}~{\rm MeV},\\
\Gamma_R &=& 144\,(169)_{\rm{Stat.}}(181)_{\rm{Total}}~{\rm MeV}.
\end{eqnarray*}

\section{Two-Pole Nature of Scattering near the \boldmath $\Lambda(1405)$}

\begin{figure}[t]
\begin{center}
\includegraphics[width=0.53\columnwidth]{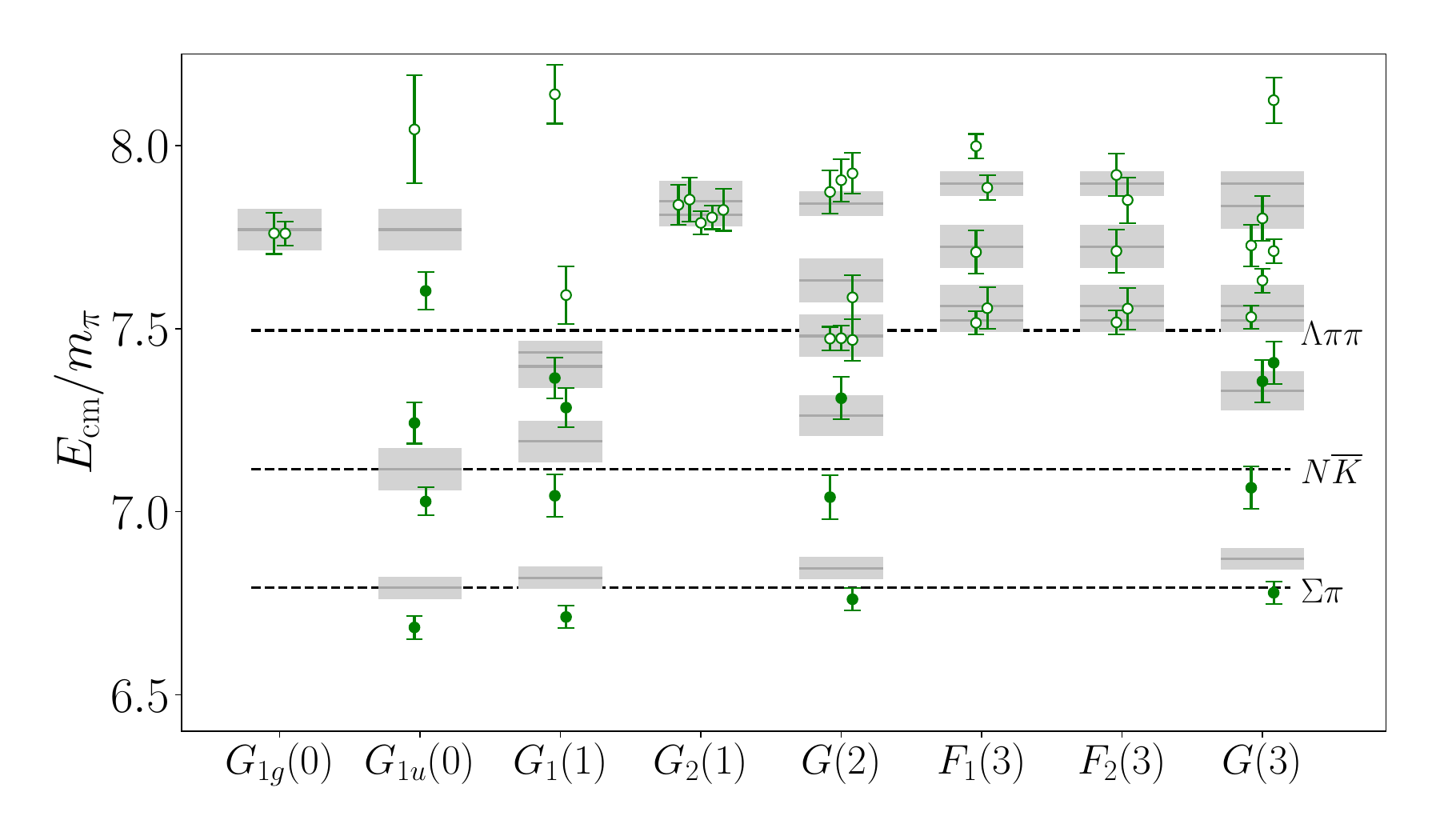}
\includegraphics[width=0.43\columnwidth]{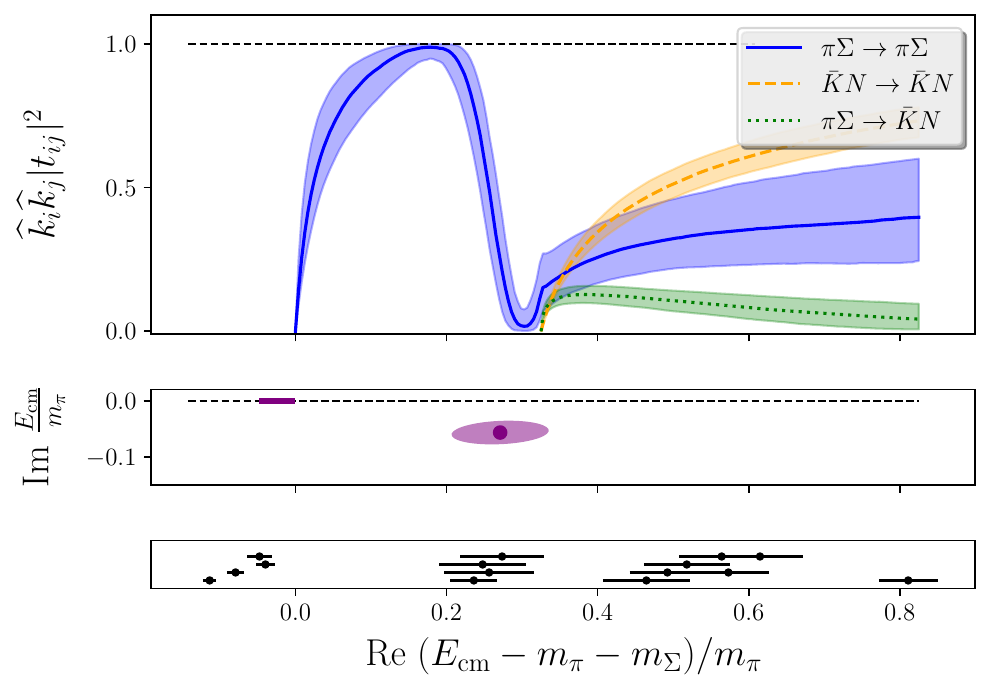}
\includegraphics[width=0.48\columnwidth]{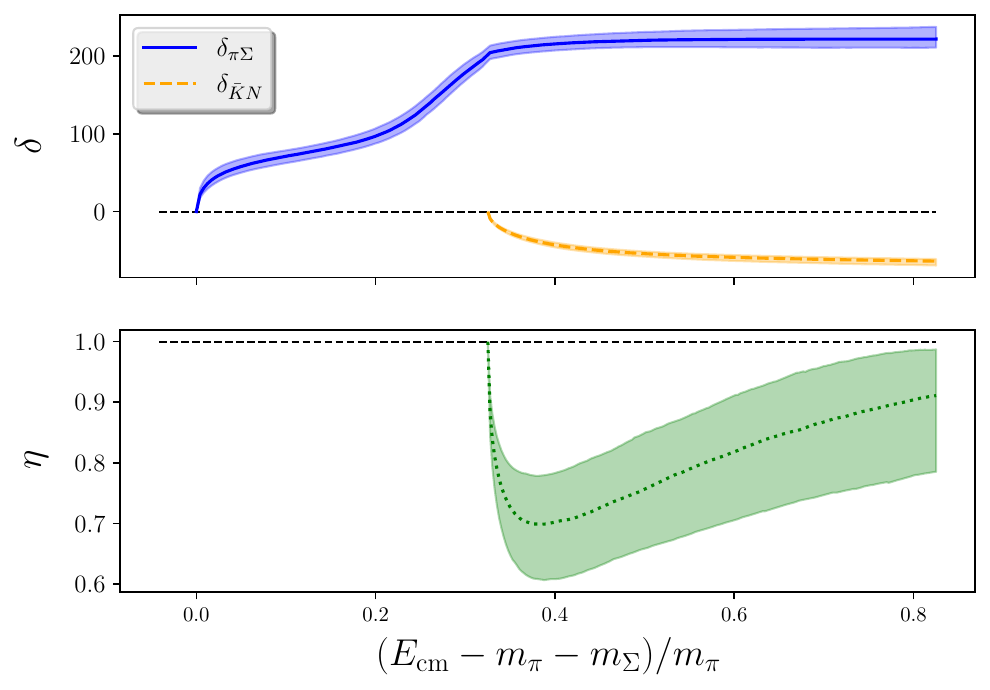}\hspace*{15mm}
\includegraphics[width=0.35\textwidth]{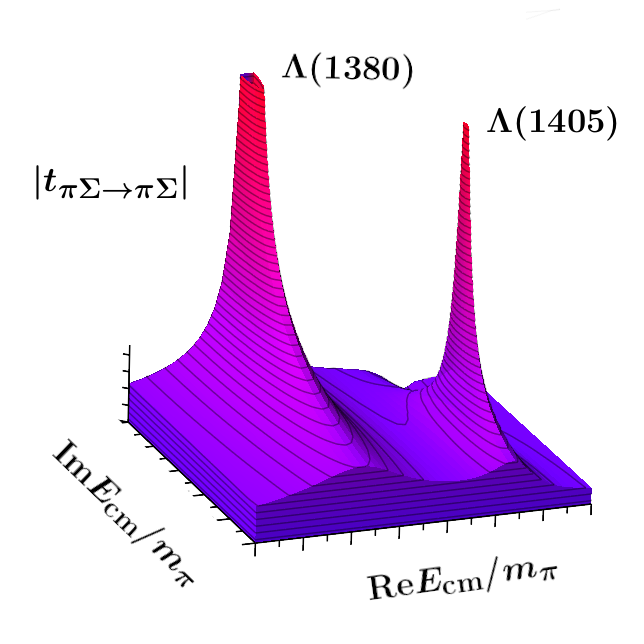}
\end{center}
\vspace*{-5mm}
\caption{(Top left) Finite volume energy spectrum involving interacting $\Sigma\pi$ and 
$N \overline{K}$ states as ratios over the pion mass
from Refs.~\cite{BaryonScatteringBaSc:2023zvt,BaryonScatteringBaSc:2023ori}.  
Green symbols are our results,
gray bands show non-interacting energies.  Labels on the horizontal axis show the irreps 
$\Lambda(\bm{d}^2)$ for lab-frame total momenta $\bm{P}=(2\pi/L)\bm{d}$, where $\bm{d}$ 
is a three-vector of integers and the lattice spatial volume is $L^3$.
(Top right) Upper panel shows the isoscalar, strangeness $-1$, $i\rightarrow j$
transition amplitudes squared for $i,j=\Sigma\pi,N\overline{K}$;
 middle panel shows positions of the $S$-matrix poles
in the complex center-of-mass energy plane on the sheet closest to the physical one;
bottom panel shows the finite-volume energies used in the fit.
(Bottom left) Inelasticity $\eta$ and phase shifts $\delta_{\pi\Sigma}$ and 
$\delta_{\overline{K}N}$.  (Bottom right) Three-dimensional plot of the 
$\Sigma\pi\rightarrow\Sigma\pi$ transition amplitude magnitude showing the two poles.
\label{fig:lambda}}
\end{figure}

An interesting energy region to probe for resonances having nonzero
strangeness is in the vicinity of the puzzling $\Lambda(1405)$ resonance.
Our recent study of $\Sigma\pi$ and $N\bar{K}$ scattering in the $\Lambda(1405)$
energy region was presented in 
Refs.~\cite{BaryonScatteringBaSc:2023zvt,BaryonScatteringBaSc:2023ori}.
Our results, shown in Fig.~\ref{fig:lambda}, were obtained using the CLS D200 ensemble
with $m_\pi\sim 200~{\rm MeV}$.
This was the first lattice QCD study of this system to include both single-hadron and
all needed two-hadron operators to carry out a full coupled-channel analysis.

The energies of the finite-volume stationary states we obtained are shown
in the upper left hand plot of Fig.~\ref{fig:lambda}. An effective range
expansion with $L_{\rm max} = 0$ of the form
\beqs
  \frac{E_{\rm cm}}{M_\pi} \tilde{K}_{ij} &=& A_{ij} + B_{ij} \Delta_{\pi \Sigma},
\\ \Delta_{\pi \Sigma} &=& (E_{\rm cm}^2 - (M_\pi+M_\Sigma)^2)/(M_\pi+M_\Sigma)^2,
 \label{eq:Kfitone}
\eeqs
where $A_{ij}$ and $B_{ij}$ are symmetric and real coefficients with $i$ and $j$ 
denoting either of the two scattering channels, provided the best description of 
the data, but several other parametrizations were
also used, including an ERE for $\tilde{K}^{-1}$, the form above with
the outer factor of $E_{\rm cm}$ removed, and a Blatt-Biedenharn form.
The fit with the lowest AIC value is a four-parameter 
fit of the form of Eq.~(\ref{eq:Kfitone}), and the best-fit parameters values are
\beq
\begin{array}{rcl@{\qquad}rcl}
    A_{00} &=& 4.2(1.8), & A_{11}&=&-10.4(1.1), \\ 
    A_{01} &=&10.4(1.6), & B_{01}&=&-30(18), 
\end{array}
\label{eq:bestfit}
\eeq
with fixed $B_{00}=B_{11}=0$ and $\chi^{2}=11.17$ for 11 degrees of freedom. This fit 
is shown in Fig.~\ref{fig:lambda}.  All statistical uncertainties and correlations 
for this fit are taken into account using the bootstrap method with $800$ samples.

To study the scattering amplitudes, we define a quantity 
$t_{ij}^{(J^P)}(E_{\rm cm})$ which is proportional 
to the scattering transition amplitude and is related to $\widetilde{K}$ by
\begin{equation}
t^{-1} = \widetilde{K}^{-1} -i \widehat k, \quad 
\label{eq:amplitude}
\end{equation}
where $\widehat k = {\rm diag }(k_{\pi \Sigma}, k_{\bar{K}N})$, with
\begin{eqnarray}
    k_{\pi\Sigma}^2=\frac{1}{4E^{2}_{\rm cm }}\lambda_K(E^{2}_{\rm cm}, m_{\pi}^{2}, m_{\Sigma}^{2}),\\
    k_{\bar{K}N}^2=\frac{1}{4E^{2}_{\rm cm }}\lambda_K(E^{2}_{\rm cm}, m_{\bar{K}}^{2}, m_{N}^{2}),
\end{eqnarray}
and $\lambda_K$ is the K\"all\'en function\cite{kallen1964}
\beq
  \lambda_K(x,y,z)=x^2+y^2+z^2-2xy-2xz-2yz.
\eeq
Results for the scattering transition amplitudes are shown in the upper right
panel of Fig.~\ref{fig:lambda}, and the pole locations for each of
the fits are shown in the lower right panel in Fig.~\ref{fig:lambda}. 
Our fits to the transition amplitudes revealed a two-pole structure, with
locations
\beqs
  E_1&=&1395(9)(2)(16)~{\rm MeV},\\
  E_2&=&[1456(14)(2)(16)-i\, 11.7(4.3)(4)(0.1)]~{\rm MeV},
\eeqs
with the first uncertainty being statistical, the second coming from our different
parametrizations of the amplitudes, and the third arising from scale setting.
A virtual bound state below the $\Sigma\pi$ threshold was found, as well as a
resonance pole below the $N\overline{K}$ threshold.  
Fit forms with just one pole were tried and all were strongly disfavored.  
The two-pole structure in
the $\Lambda(1405)$ region was first suggested in Ref.~\cite{Oller:2000fj}.
The scattering phase shifts $\delta_i$ and the inelasticity $\eta$ are related 
to $t$ by
\beqs
t_{00}  &=&\frac{\eta e^{2 i \delta_{\pi\Sigma}}-1}{2 i \widehat k_{\pi\Sigma}}, \\
t_{11}  &=&\frac{\eta e^{2 i \delta_{\bar{K}N}}-1}{2 i \widehat k_{\bar{K}N}} , \\
t_{01}  &=&\frac{\sqrt{1-\eta^2} e^{i\left(\delta_{\pi\Sigma}+\delta_{\bar{K}N}\right)}}{2 
  \sqrt{\widehat k_{\pi\Sigma}\widehat k_{\bar{K}N}}},
\label{eq:inelas}
\eeqs
where the indices indicate the flavor channel: $0$ for $\pi\Sigma$ and 1 for 
$\bar{K}N$. The results are shown in the lower left panel of Fig.~\ref{fig:lambda}. 

The analysis above only includes $S$-waves and the $J^P=1/2^-$ amplitudes.
The energies in nonzero momentum frames, such as the $G_1(1)$, $G(2)$, and $G(3)$
irreps, receive contamination from $P$-waves.  To assess the importance of
such contributions, additional fits with a very simple parametrization
for the higher partial waves were done.  The effects of the higher partial
waves were found to be negligible for energies below the $\Lambda\pi\pi$ threshold.

\begin{figure}
\begin{center}
\includegraphics[width=3.7in]{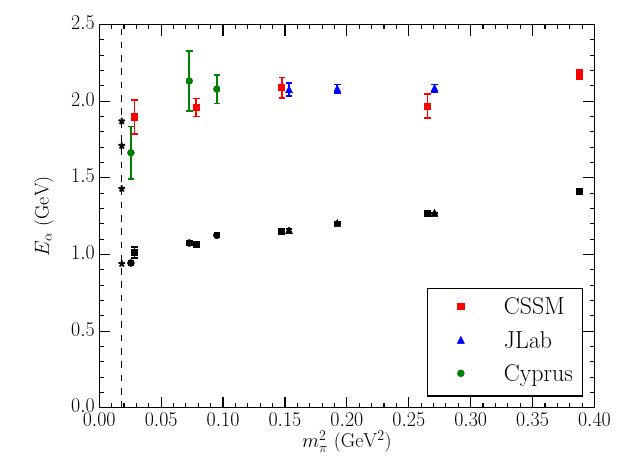}\\
\includegraphics[width=4.5in]{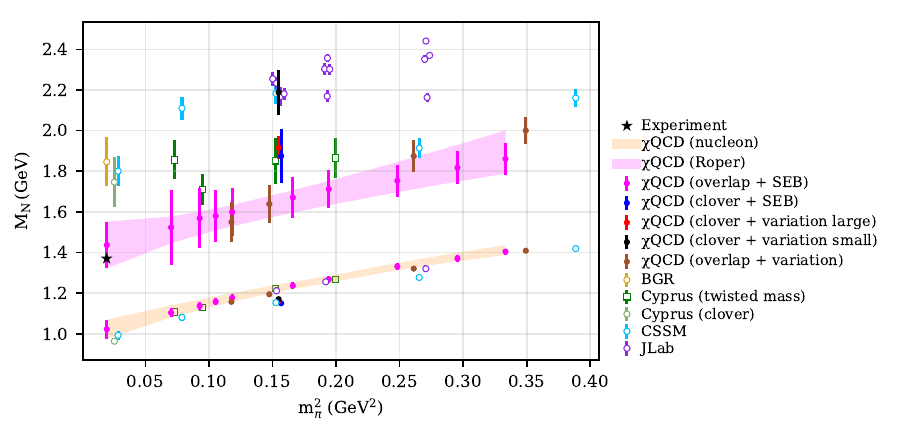}
\end{center}
\caption{(Top) Current status of positive-parity excitations of the proton
from Ref.~\cite{LeinweberNSTAR2024} which compares results from 
Refs.~\cite{Alexandrou:2014mka,Edwards:2011jj,Liu:2016uzk}.  Such studies
which only use three-quark operators miss the Roper and instead capture
a higher lying excitation. (Bottom) Results for the positive-parity
excitation spectrum of the proton from Ref.~\cite{xQCD:2019jke} are shown.
A large number of differently-smeared three-quark operators, combined with
a domain-wall fermion sea, overlap valence fermions, and a sequential Bayesian
analysis method, seems to capture the Roper (magenta band), but with rather
large uncertainties.
\label{fig:roper}}
\end{figure}

\section{The Roper resonance}

The first excitation of the proton, known as the Roper resonance, is an important
resonance.  Experimentally, it is a 4-star resonance $N(1440)$ with 
$I(J^P)=\frac{1}{2}(\frac{1}{2}^+)$ and a width in the range $250-450~{\rm MeV}$.
It is a notoriously difficult resonance to study in lattice QCD.  Local three-quark
operators have difficulty capturing the Roper level near 1.4~GeV and instead 
create a state with an energy much higher near 2.0~GeV.  This fact is illustrated
in the upper plot of Fig.~\ref{fig:roper} which shows energy extractions for the
proton and its first excitation from three lattice QCD studies.

The Roper resonance was studied in Ref.~\cite{xQCD:2019jke} using only a selection of 
three-quark operators with domain-wall fermions in the sea and overlap fermions
for the valence quarks.  Their results, shown in the lower panel in 
Fig.~\ref{fig:roper}, are obtained employing a large basis of three-quark operators
with different quark-field smearings and a sequential
empirical Bayesian analysis method.  The Roper mass does seem to be captured,
but with very large uncertainties.

It is evident that a definitive study of the Roper resonance needs 
multi-hadron operators involving $N\pi$, $N\sigma$, $\Delta\pi$ operators,
as well as $N\pi\pi$ operators.  Large volumes will be needed, as well as
a three-particle amplitude analysis, which has become available only 
recently\cite{Hansen:2025oag}.

\section{Conclusion}

Recent innovations in lattice QCD methods, such as the stochastic LapH method
and distillation, have facilitated reliable determinations of energies involving
multi-hadron states.
Large numbers of excited-state energy levels can now be estimated, allowing
scattering phase shifts to be computed and hadron resonance properties,
such as masses and decay widths, to be determined.  Our
recent results for the $\Delta$\ and $\Lambda(1405)$ resonances from
lattice QCD were highlighted.  The famous Roper resonance is still a challenge,
but future studies involving three-particle operators may finally
shed light on this elusive hadron.  

\section*{Acknowledgments}

The works described here were done in collaboration with John Bulava, 
B\'arbara Cid-Mora, Andrew D. Hanlon, Ben H\"orz, Daniel Mohler, 
Joseph Moscoso, Amy Nicholson, Fernando Romero-L\'opez, Sarah Skinner, 
Pavlos Vranas, and Andr\'e Walker-Loud.
The author gratefully acknowledges support from 
the U.S.~National Science Foundation (NSF)
under awards PHY-2209167 and PHY-2514831.
Computations were carried out on Frontera\cite{stanzione2020frontera} 
at the Texas Advanced Computing
Center (TACC) under award PHY20009, and at the 
National Energy Research Scientific
Computing Center (NERSC), a U.S.~Department of Energy (DOE)
Office of Science User 
Facility located at Lawrence Berkeley National Laboratory (LBNL),
operated under 
Contract No. DE-AC02-05CH11231 using NERSC awards NP-ERCAP0005287, 
NP-ERCAP0010836 and NP-ERCAP0015497.

\bibliographystyle{JHEP}
\bibliography{references}

\end{document}